\documentclass[%
 reprint,
superscriptaddress,
 amsmath,amssymb,
prb,
]{revtex4-2}

\usepackage{graphicx}
\usepackage{dcolumn}
\usepackage{bm}

\begin{document}

\title{Probing the Semiconductor-to-Dirac Semimetal Transition in Na–Sb–Bi Alloys with x-ray Compton Scattering}


\author{Aki Pulkkinen}
\email{apulkkin@ntc.zcu.cz}
\affiliation{%
New Technologies-Research Centre, University of West Bohemia, 30100 Plzeň, Czech Republic}%
\author{Veenavee Nipunika Kothalawala}
\affiliation{Department of Physics, School of Engineering Science, LUT University, FI-53851 Lappeenranta, Finland}
\author{Kosuke Suzuki}%
\affiliation{Graduate School of Science and Technology, Gunma University, Kiryu, Gunma 376-8515, Japan}
\author{Bernardo Barbiellini}
\affiliation{Department of Physics, School of Engineering Science, LUT University, FI-53851 Lappeenranta, Finland}
\affiliation{Department of Physics, Northeastern University, Boston, Massachusetts  02115, USA}
\affiliation{Quantum Materials and Sensing Institute, Northeastern University, Burlington, MA 01803, USA}
\author{Johannes Nokelainen}
\affiliation{Department of Physics, School of Engineering Science, LUT University, FI-53851 Lappeenranta, Finland}
\affiliation{Department of Physics, Northeastern University, Boston, Massachusetts  02115, USA}
\affiliation{Quantum Materials and Sensing Institute, Northeastern University, Burlington, MA 01803, USA}
\author{Wei-Chi Chiu}
\affiliation{Department of Physics, Northeastern University, Boston, Massachusetts  02115, USA}
\affiliation{Quantum Materials and Sensing Institute, Northeastern University, Burlington, MA 01803, USA}
\author{Bahadur Singh}
\affiliation{Department of Condensed Matter Physics and Materials Science, Tata Institute of Fundamental Research, Mumbai 400005, India}
\author{Hsin Lin}
\affiliation{Institute of Physics, Academia Sinica, Taipei 11529, Taiwan}
\author{Alok K. Pandey}
\affiliation{Department of Chemistry and Life Science, Yokohama National University, 79-5 Tokiwadai, Hodogaya-ku, Yokohama, Kanagawa 240-8501, Japan}
\author{Naoaki Yabuuchi}
\affiliation{Department of Chemistry and Life Science, Yokohama National University, 79-5 Tokiwadai, Hodogaya-ku, Yokohama, Kanagawa 240-8501, Japan}
\author{Naruki Tsuji}
\affiliation{Japan Synchrotron Radiation Research Institute (JASRI), Sayo, Hyogo 679-5198, Japan}
\author{Yoshiharu Sakurai}
\affiliation{Japan Synchrotron Radiation Research Institute (JASRI), Sayo, Hyogo 679-5198, Japan}
\author{Hiroshi Sakurai}
\affiliation{Graduate School of Science and Technology, Gunma University, Kiryu, Gunma 376-8515, Japan}
\author{Ján Minár}
\affiliation{%
New Technologies Research Centre, 30100 Pilsen, Czech Republic}
\author{Arun Bansil}
\affiliation{Department of Physics, Northeastern University, Boston, Massachusetts  02115, USA}
\affiliation{Quantum Materials and Sensing Institute, Northeastern University, Burlington, MA 01803, USA}

\date{\today}

\begin{abstract}
{We discuss electron redistribution during the semiconductor-to-Dirac semimetal transition in Na-Sb-Bi alloys using x-ray Compton scattering experiments combined with first-principles electronic structure modeling. A robust signature of the semiconductor-to-Dirac semimetal transition is identified in the spherically averaged Compton profile. We demonstrate how the number of electrons involved in this transition can be estimated to provide a novel descriptor for quantifying the strength of spin-orbit coupling responsible for driving the transition. The associated theoretical deviation of the Born charge of Na in Na$_3$Bi from the expected ionic charge of +1 is found to be consistent with the corresponding experimental value of about 10\%. Our study also shows the sensitivity of the Compton scattering technique toward capturing the spillover of Bi 6p relativistic states onto Na sites.}
\end{abstract}

\maketitle
\section{introduction}
Na$_3$Sb has attracted considerable interest due to its unique electronic properties, particularly in the context of topological transitions in the Na-Sb-Bi alloys \cite{narayan2014,chiu2020}. It contains earth-abundant antimony (Sb), a cost-effective material that is also used in Na-ion battery anodes \cite{darwiche2012,darwiche2018,he2018Sb}. Na$_3$Sb involves an ionic bond between the positively charged Na$^+$ and negatively charged Sb$^{3-}$ ions, resulting in a fully occupied 5p$^6$ Sb shell that introduces an energy gap which reduces conductivity.

A rigorous definition of oxidation states of ions in solids based on Wannier centers \cite{jiang2012} yields an integer value of +1 for the oxidation state of Na in Na$_3$Sb. 
{However, such well-defined oxidation states that are typical of ionic solids and electrolytes \cite{Pegolo2020,resta2021} need not hold when Sb is replaced by Bi because Bi is a heavier element \cite{hsu2019} and its 6p orbitals experience significant spin-orbit coupling (SOC), which splits the p orbitals into p$_{1/2}$ (lower energy) and p$_{3/2}$ (higher energy) components \cite{pyykko1979,norrby1991}. This splitting results in a substantial difference in the associated ionic radii, with the 6p$_{1/2}$ orbital having a radius of 1.48 \AA\  and the 6p$_{3/2}$ orbital a radius of 1.67 \AA\  \cite{pyper2020}. Bi thus adopts a valence electron configuration of (6s)$^2$(6p$_{1/2}$)$^2$(6p$_{3/2}$)$^1$, where the outermost electron, which occupies the 6p$_{3/2}$ orbital, is weakly bound and destabilized by the SOC effects. For the three Na atoms to lose electrons and form a Bi$^{3-}$ ion \cite{chen2018}, a reduction in SOC would be required. At a critical SOC strength, the energy gap between the 3s band of Na and the 6p band of Bi closes to yield a topological phase transition. Na-Sb-Bi alloys thus offer unique opportunities for tuning electronic properties through the interplay of composition,
topology, and SOC. Notably, the end-compound Na$_3$Bi is a three-dimensional (3D) Dirac semimetal (DSM) \cite{Liu2014,Bansil2016,Jenkins2016,dibernardo2021}.

Analysis of electron momentum density using x-ray Compton scattering techniques \cite{Cooper2004,barbiellini2024,Kaplan2003} has emerged as a powerful tool for investigating electronic structures of wide classes of materials. Here, by combining x-ray Compton scattering measurements with fully relativistic Density Functional Theory (DFT) calculations, we unravel the role of {SOC effects} in inducing displacements in the electron momentum density and give insight into the nature of topological transitions in Na-Bi-Sb alloys. We will show how by extracting the valence momentum density of the light element Na embedded in the Sb-Bi matrix, we can estimate the electronic charge displaced (denoted by $q$) through SOC effects. $q$ describes the modulation of the momentum density and quantifies the strength of the SOC. When a semiconductor-to-DSM transition is induced by tuning the SOC strength, the ionic-like bonding of the trivial phase with localized electrons crosses over to becoming metallic with delocalized states.}

The measured spectrum in a Compton scattering experiment is the one-dimensional profile, $J(p_z)$, which represents a projection of the 3D electron momentum density $\rho(p_x,p_y,p_z)$ along the $z$-axis that is defined to lie along the direction of the scattering vector: 
\begin{equation}
J(p_z) = \int \int \rho(p_x,p_y,p_z) \, dp_x \, dp_y ~.
\label{eq1}
\end{equation}
Since our focus in this study is on spherically averaged spectra, we will hereafter replace $p_z$ by the radial distance $p$.

\section{methods}
\subsection{Sample preparation}
Na$_3$Sb sample was synthesized using a heat treatment method. Finely cut metallic sodium was first mixed with antimony powder by hand milling. The mixture was then covered by a copper foil and fired at 400\, $^{\circ}$C  for 10\,h in a furnace placed in an argon-filled glove box to obtain the alloy material. Excess (4 wt\%) metallic sodium was used to compensate for Na loss during heating. Na$_3$Bi was synthesized using a two-step heat treatment method in which the finely cut metallic sodium was thoroughly mixed with bismuth powder by hand milling. The mixture was then covered by a copper foil and fired at 100\, $^{\circ}$C for 10\,h in a furnace placed in an argon-filled atmosphere. The Na-Bi mixture was ground by hand milling and heated again for 10 hours at 400\,$^{\circ}$C in an argon atmosphere to obtain the alloy material. Here also excess (4 wt\%) metallic sodium was used to compensate for Na loss during heating. {A sodiated sample with equal concentrations of Sb and Bi was produced
following the same procedure, using an excess of 15 wt\% metallic sodium. The final stoichiometry was determined to be  Na$_{3.3}$Sb$_{0.5}$Bi$_{0.5}$ through precise Compton core electron analysis as described 
in Ref. \cite{kothalawala2024b}}. 

\subsection{Experiment}
{Compton profiles of the alloys, as well as pristine Sb and Bi, were measured at room temperature using the high-energy inelastic x-ray scattering beamline 08W at SPring-8, Japan}. The incident x-ray energy was 182.6~keV. The scattering angle was fixed at 178$^{\circ}$ and the energy spectrum of scattered x-rays was measured using a pure Ge solid-state detector with 10 elements. The samples were packed in a 3\,mm deep hole of 2\,mm diameter in the sample holder. The single-phase hexagonal structure of both Na$_3$Bi and Na$_3$Sb samples was confirmed by XRD measurements. Compton profiles were obtained by summing up the spectra obtained from various detectors after applying corrections for scattering angle, multiple scattering, and detector efficiency. {Multiple scattering corrections are described in references \cite{Cooper2004,robats2020} but can be neglected in powder samples.}  The overall momentum resolution of the measurements is 0.5\,a.u.

\subsection{Computations}
\begin{figure}
\includegraphics[width=\columnwidth]{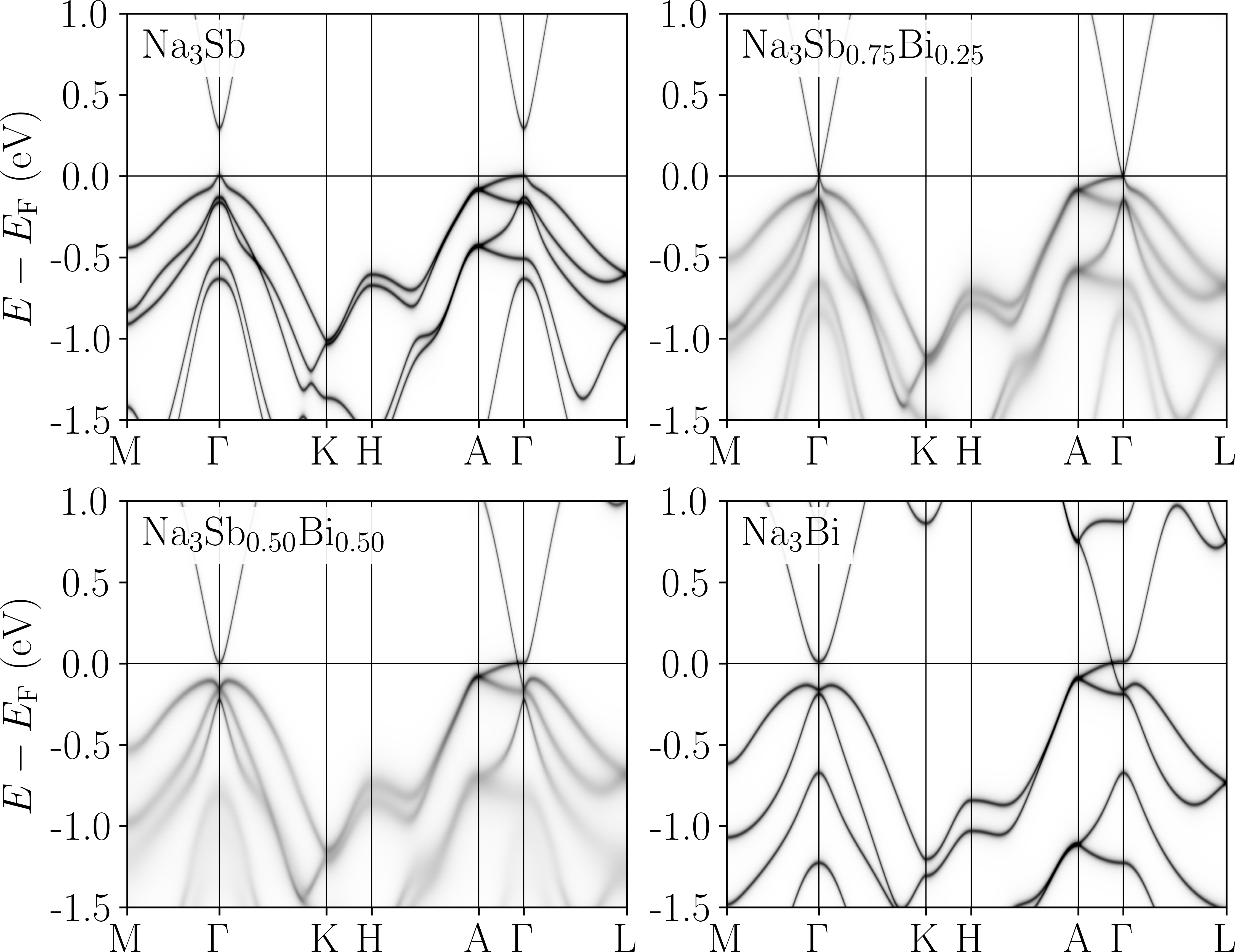}
\caption{\label{fig:BSF}Bloch spectral functions of Na$_3$Sb, Na$_3$Sb$_{0.75}$Bi$_{0.25}$, Na$_3$Sb$_{0.50}$Bi$_{0.50}$ and Na$_3$Bi.}
\label{fig:bands}
\end{figure}

The Bloch spectral functions and Compton profiles were calculated using the fully relativistic, full-potential Korringa-Kohn-Rostoker method as implemented in the SPRKKR package~\cite{Ebert_2011} with a basis set truncated at $l_{\rm max} = 3$ within the coherent potential approximation (CPA)~\cite{Bansil1981,Bansil1999,Benea_2006,Benea_2018}. 
Exchange-correlation (xc) effects were treated at the level of the generalized gradient approximation (GGA) using the Perdew-Burke-Ernzerhof (PBE) functional~\cite{PhysRevLett.77.3865}. The results, shown in Fig.~\ref{fig:bands}, are consistent with both the CPA calculations of Narayan {\it et al.} \cite{narayan2014}.
However, the present DFT results slightly differ from those based on the virtual crystal approximation by Chiu et al. \cite{chiu2020}. In Fig. 4(b) of Ref.~\cite{chiu2020}, the gap at the $\Gamma$ point remains open in Na$_3$(Sb$_{1-x}$Bi$_x$) until $x \textgreater  0.6$, while in Fig.~\ref{fig:bands} of this work, the gap closes already at $x = 0.25$.

For comparison with experimental Compton profiles from powder samples, the spherically averaged Compton profiles were calculated as a weighted average over a set of five special directions \cite{bansil1975} appropriate for the hcp lattice~\cite{Kontrym-Sznajd_2015}. Note that although Na$_3$Bi and Na$_3$Sb share the same hcp structure (space group no. 194), their $c/a$ ratio is $8.5 \%$ larger than the ideal hcp structure. Further technical details about the calculated and measured spherical Compton profiles are available in the Supplemental Material.

\section{Results}

\begin{figure}
\includegraphics[width=\columnwidth]{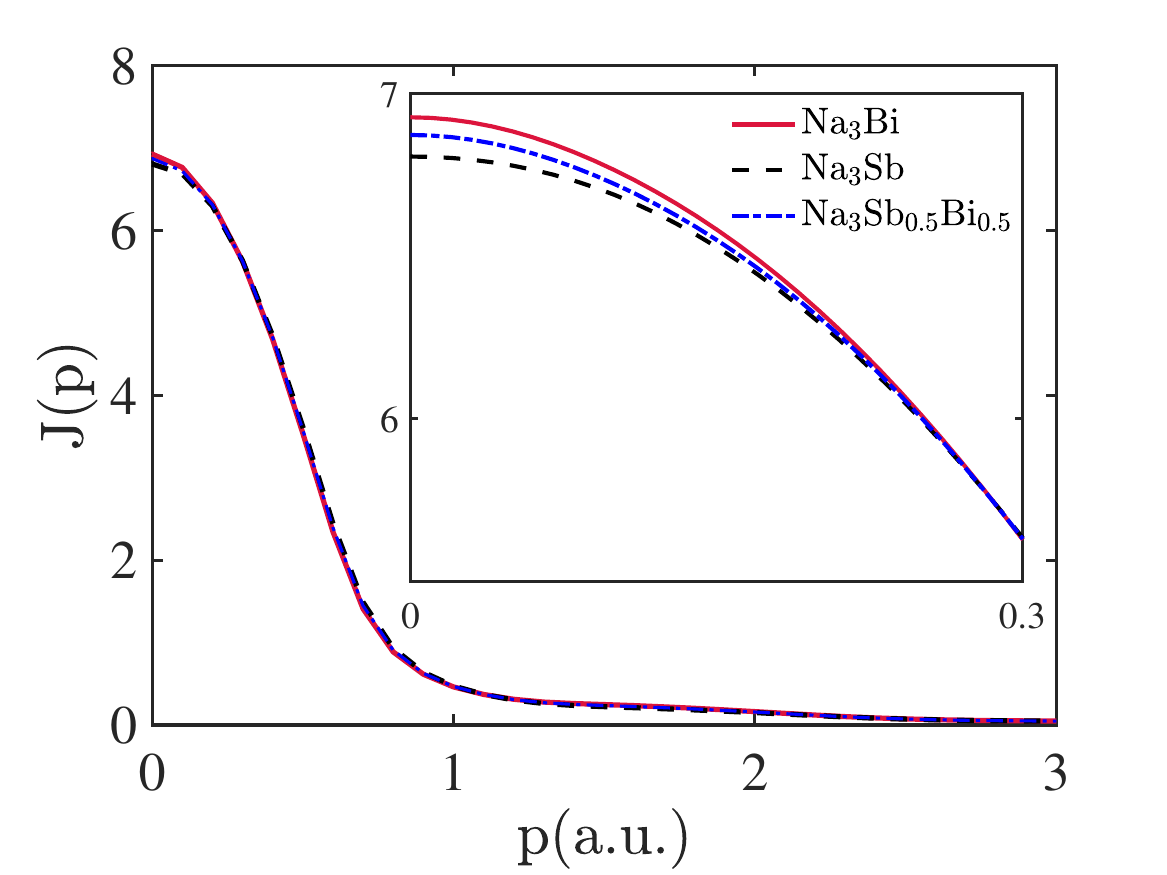}
\caption{Spherically-averaged computed valence Compton profiles, $J(p)$, for Na$_3$Bi (solid red lines) and Na$_3$Sb (dashed black lines). Inset shows a magnified view to highlight the differences between the two profiles around the peak position at $p=0$. 
The models are convolved with the 0.5 a.u. momentum resolution of the measurements.}
\label{fig:compton}
\end{figure}

The DFT based valence Compton profiles shown in Fig.~\ref{fig:compton} include contributions from Na 3s, Bi 6s and Bi 6p electrons in Na$_3$Bi, and Na 3s, Sb 5s and Sb 5p electrons in Na$_3$Sb. The profiles are normalized such that the total area under each profile over the $-10$ a.u. to $10$ a.u. range equals the number of valence electrons in the unit cell. 
Na$_3$Bi exhibits modifications in the low-momentum region, confirming the delocalization of Bi 6p electrons. Relativistic SOC effects play a key role in the splitting and rearrangement of electronic levels in heavy elements such as bismuth, influencing the spatial electronic distributions. In Na$_3$Bi, this results in changes in the momentum distribution of 6p electrons, and increased delocalization compared to the non-relativistic case. Understanding these relativistic effects is important to elucidate the unique electronic properties of Na$_3$Bi and its DSM state.

\begin{figure}[h]
\includegraphics[width=\columnwidth] {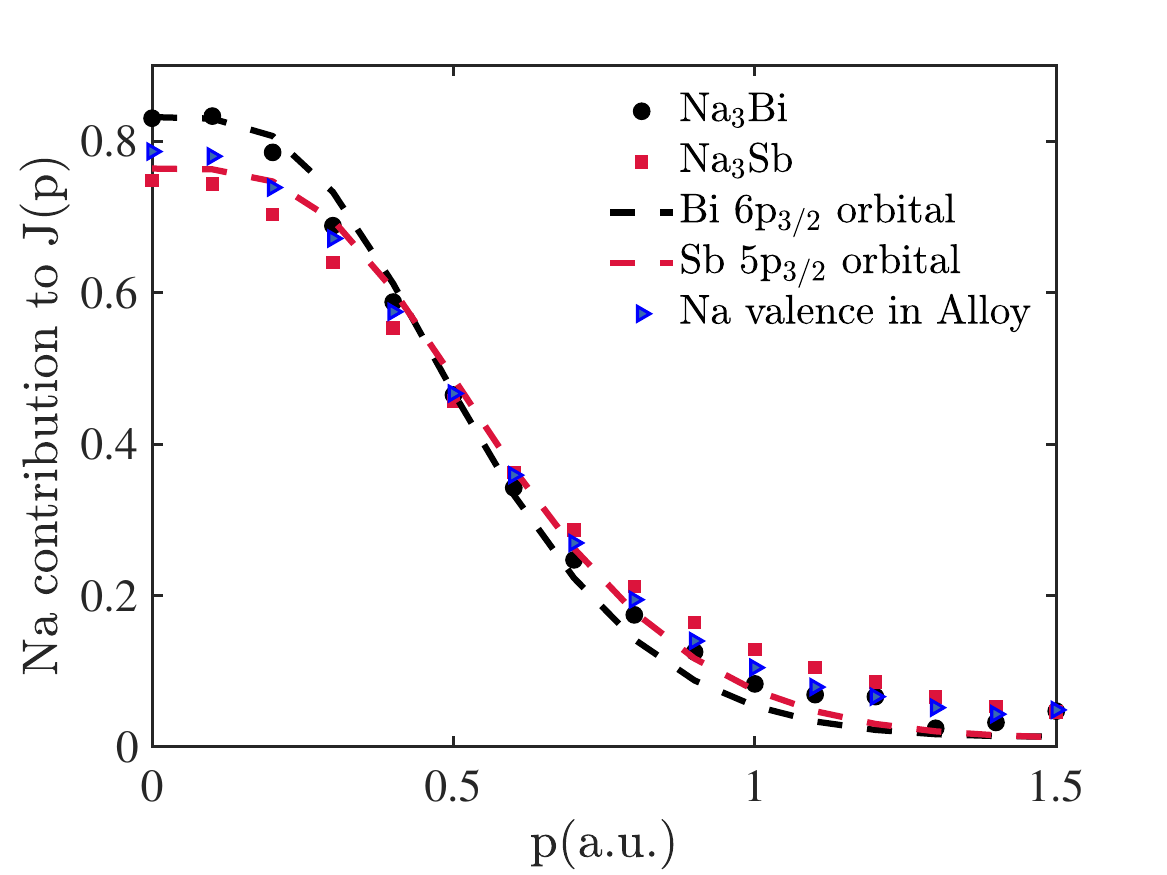}
\caption{{Na contribution to the experimental valence Compton profiles of Na$_3$Bi (black markers), Na$_3$Sb (red markers) and Na$_{3.3}$Sb$_{0.5}$Bi$_{0.5}$ (blue triangle markers). Size of the markers represents experimental error bars. The black dashed line gives the theoretical relativistic profile for the atomic Bi 6p$_{3/2}$ orbital while the red dashed line gives the profile for the (less relativistic) Sb 5p$_{3/2}$ orbital. Theory profiles have been convolved with the momentum resolution of 0.5 a.u. of the measurements.} }
\label{fig:orbitals}
\end{figure}

{To obtain a handle on the electron redistribution during the gap closing, we used the robust approach of Kothalawala {\it et al.} \cite{kothalawala2024}, which allows to isolate the contribution of the Na valence electron using the following steps.} (1) The sodium (Na) contribution is isolated taking the profile differences [${\rm Na}_3{\rm Bi} - {\rm Bi}$] and [${\rm Na}_3{\rm Sb~} - {\rm Sb}$]. {In the case of Na$_{3.3}$Sb$_{0.5}$Bi$_{0.5}$, the difference is taken using the averaged profile of Sb and Bi.} The experimental profiles for pure Bi and pure Sb are used in these subtractions. Taking profile differences, we remove a variety of irrelevant background contributions \cite{kothalawala2024}. (2) The Na core electron contribution is removed using the Na core electron Compton profile obtained from relativistic Hartree-Fock calculations\cite{biggs1975}. (3) The valence momentum range is limited to [$-p_{\text{max}}$, $p_{\text{max}}$], where $p_{\text{max}}=1.5$ a.u. is a reasonable momentum cut-off to isolate the {s$-$p} valence contribution. Finally, the profiles are renormalized to correspond to one electron. This adjusts the profiles to a common scale and allows comparison of shapes. Fig~\ref{fig:orbitals} shows the Na contribution to the Compton profiles in the Sb and Bi matrices obtained by following the preceding steps. These results indicate that the Na 3s electron is donated to bands characterized by the relativistic Bi 6p$_{3/2}$ orbital with 1.67 \AA\ \cite{pyper2020} radius and the less relativistic Sb 5p$_{3/2}$ orbital with a smaller radius. This atomic model thus provides a reasonable understanding of the electronic interactions within the Na-Sb-Bi alloys and serves as a good basis for our analysis.
{Interestingly, since Na$_3$Bi is a DSM and Na$_3$Sb is a normal insulator, the Compton profile measurements show that $J(p)$ of the intermediate alloy falls between those of the two end compounds, as seen in Fig.~\ref{fig:orbitals}. This raises the question whether the alloy is topologically trivial. In Fig.~\ref{fig:orbitals}, the Na contribution to $J(p)$ (blue triangle markers) appears to align more closely to the insulating Na$_3$Sb phase. However, this seems in contradiction with the DFT results in Fig.~\ref{fig:bands}, which indicate a Fermi level crossing near the $\Gamma$ point for Na$_{3}$Sb$_{0.5}$Bi$_{0.5}$. This small discrepancy can be resolved by incorporating a meta-GGA correction, as shown in Ref.~\cite{chiu2020}, where the intermediate alloy becomes topologically trivial.}

\begin{figure}[h]
\includegraphics[width=\columnwidth]{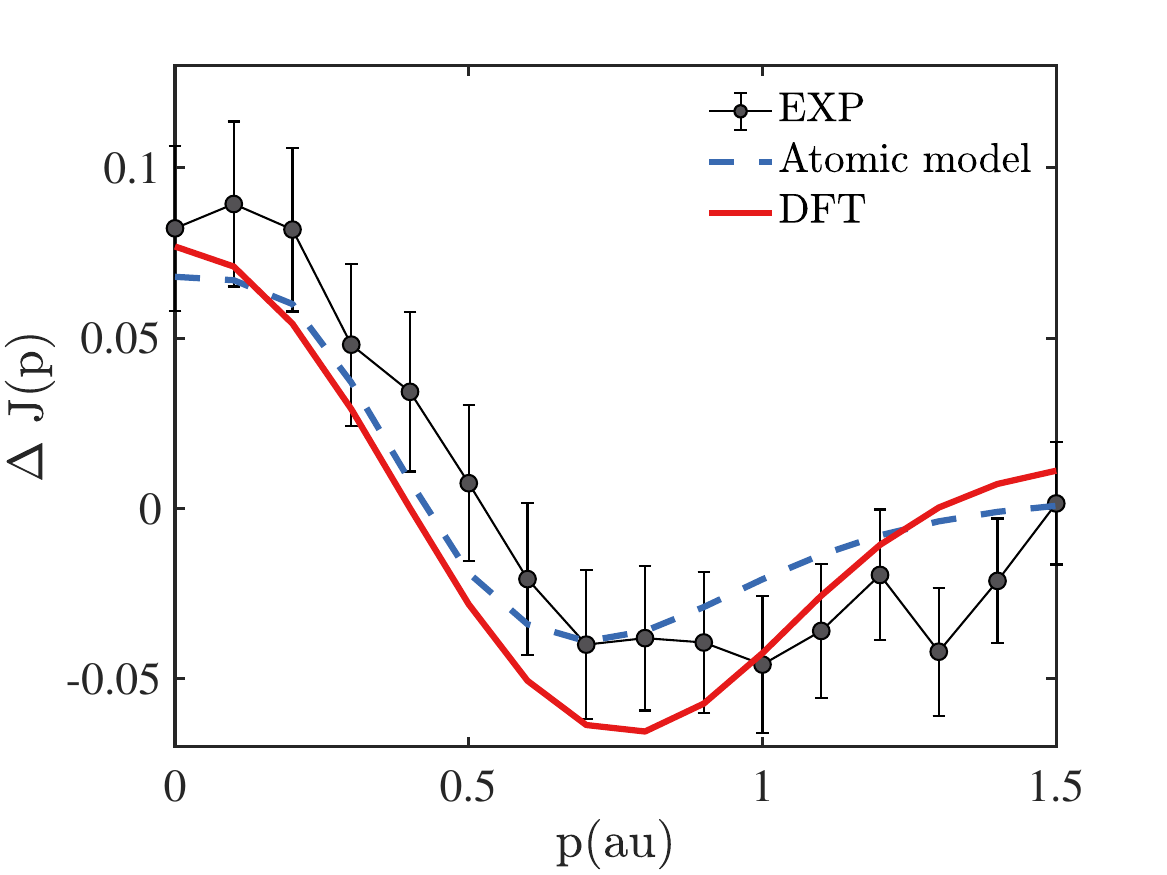}
\caption{\label{fig:topological profile} Theoretical and experimental profiles, $\Delta J(p)$. Experimental $\Delta J(p)$ (filled black dots joined by solid black line to guide the eye) is obtained by taking the difference between the profiles marked as black and red in Fig~\ref{fig:orbitals}. Atomic model (blue dashed line) refers to the difference between the theoretical atomic profile of the relativistic Bi 6 $p_{3/2}$ orbital and the (less relativistic) 5p Sb $p_{3/2}$ orbital in Fig. 3. DFT curve (solid red line) is obtained by taking the difference $\Delta J(p)=J_{{\rm Na}_3{\rm Bi}-{\rm Bi}}(p)- 
J_{{\rm Na}_3{\rm Sb}-{\rm Sb}}(p)$. The theory curves (atomic model and DFT) are convolved with the 0.5 a.u. experimental momentum resolution. }
\label{fig:topo}
\end{figure}

{We are now in a position to identify the electron redistribution $\Delta J(p)$ in the Compton profile by taking the difference between the Na contributions to the Na$_3$Bi and Na$_3$Sb (filled black circles and red squares in Fig~\ref{fig:orbitals}):
\begin{equation}
\Delta J(p)=J_{\mbox{Na$_3$Bi-Bi}}(p)- 
J_{\mbox{Na$_3$Sb-Sb}}(p)~.
\end{equation}
The results shown in Fig.~\ref{fig:topo} are similar to those based on the atomic model (within experimental error bars). To obtain $\Delta J(p)$ from DFT-based profiles, we follow the approach of Kothalawala {\em et al.} \cite{kothalawala2024} along the steps discussed above. We have checked that the fully relativistic SPR-KKR Compton profile in Figure 2 accurately matches the spherically averaged profile (with SOC) used in Ref.~\cite{kothalawala2024}.} {The charge $q$ displaced by the transition can now be quantified by calculating the area under the DFT-derived topological Compton profile $\Delta J_{\mbox{DFT}}(p)$ (without convolving with experimental resolution) as:}
\begin{equation}
q = \frac{1}{2} \int_{-p_{\text{max}}}^{p_{\text{max}}} dp \left| \Delta J_{\mbox{DFT}}(p) \right|~.
\label{eq9}
\end{equation}
The value of $q$ obtained from our analysis is 10.4\% of an electron. After convolving the profile with the experimental resolution, $q$ is reduced to 5\%. Interestingly, the Born effective charge of Na in Na$_3$Bi is 0.9~\cite{dong2019}, deviating by 10\% from the nominal ionic charge of +1. This deviation is remarkably close to the DFT-based value of $q$. It is worth noting that Born effective charges, which describe the response of electronic polarization to atomic displacements in both metals and insulators, are linked to the Berry curvature—a topological property of Bloch wavefunctions~\cite{resta2023}. By performing linear extrapolations using the present DFT band structure and the results reported by Chiu et al.~\cite{chiu2020}, we estimate that the band gap closes when $q$ lies between 2.5\% and 6\% of an electron. This uncertainty arises primarily from correlation effects. More accurate predictions may be achieved by employing advanced exchange-correlation functionals in future calculations.

\section{Conclusion}

Compton scattering is a powerful, bulk-sensitive probe for investigating electron redistribution during gap-closing transitions. By analyzing changes in the Compton profile, this technique quantifies the number of electrons involved in metal-insulator transitions and provides insight into their orbital character. Unlike surface-sensitive methods, Compton scattering reveals intrinsic bulk electronic behavior, making it particularly valuable for studying systems such as vanadium oxides \cite{ruotsalainen2018}, manganites \cite{barbiellini2009}, and the present Na-Bi-Sn alloys, where such transitions are prevalent. The integration of difference profiles offers a direct measure of electron participation, while first-principles calculations enable detailed interpretation of the associated orbital contributions. This approach complements existing spectroscopic techniques and deepens our understanding of the microscopic mechanisms underlying gap-closing phenomena.
In the present case, our DFT-based analysis shows that relativistic effects lead to an extension of the Bi 6p states onto the Na sites, driving the formation of the topological Dirac semimetal (DSM) phase. From the analysis of the profile $\Delta J(p)$, we estimate that approximately 0.1 electrons per Na atom are involved in the electron redistribution triggered by the substitution of Sb with Bi. These findings show the utility of Compton scattering as a robust tool for understanding the DSM state in Na$_3$Bi.

\begin{acknowledgments}
We are grateful to Dr. Christopher Lane for important discussions. B.B. is grateful to Riccardo Comin for valuable discussions and generous hospitality at MIT. This research was supported by the INERCOM platform at LUT University. The authors wish to acknowledge the CSC-IT Center for Science, Finland, for computational resources. J.M. and A.P. would like to thank the project Quantum materials for applications in sustainable technologies (QM4ST), funded as Project No. CZ.02.01.01/00/22{\_}008/0004572 by Programme Johannes Amos Comenius, call Excellent Research. K.S. was supported by JSPS KAKENHI Grant Nos. 19K05519, 22H02103 and 23K23371. Compton scattering experiments were performed with the approval of JASRI (Proposal Nos. 2020A1497 and 2022A1446). H.L. acknowledges the support by Academia Sinica in Taiwan under grant number AS-iMATE-113-15. The work at TIFR Mumbai was supported by the Department of Atomic Energy of the Government of India under Project No. 12-R\&D-TFR-5.10-0100 and benefited from the computational resources of TIFR Mumbai. The work at Northeastern University was supported by the U.S. Office of Naval Research grant number N00014-23-1-2330 and benefited from the resources of Northeastern University’s Advanced Scientific Computation Center, the Discovery Cluster, the Massachusetts Technology Collaborative award MTC-22032, and the Quantum Materials and Sensing Institute.
\end{acknowledgments}


\bibliography{References}

\end{document}